\documentstyle[12pt,aasms4] {article}

\begin{document}

\title{\bf Constraining The Magnetic Field in  Gamma Ray Burst Blast Waves}

\author{\it Abhas Mitra\altaffilmark{1}}

\altaffiltext{1}{Theoretical Physics Division, Bhabha Atomic Research Center,
Mumbai- 400 085, India
E-mail: nrl@magnum.barc.ernet.in, also krsrini@magnum.barc.ernet.in}

\begin{abstract}
We point out that already existing literature on relativistic
collisionless MHD shocks show that the parameter $\sigma\equiv$ upstream
proper magnetic energy density/upstream rest mass energy density, plays an
important role in determining the structure and accelerating properties of
such shocks. By adopting the value of $\sigma\approx 0.002$  which
corresponds to the relativistic shock associated with the Crab nebula, and
by using appropriate relativistic shock jump conditions, we obtain here  a
generous upper-limit on the value of (proper) magnetic field,  $B_{\rm
sh}\approx 1.5 \times 10^{-3} \eta n_1^{1/2} $G, for gamma ray burst (GRB)
blast wave.  Here, $\eta \equiv E/Mc^2$, where $E$ is the  energy and $M$
is the mass of the baryons entrained in the original fireball (FB), and
$n_1$ is the proper number density of the ambient medium. Further, we
point out that, in realistic cases, the actual value  $B_{\rm sh}$ could
be as low as $\sim 5\times 10^{-6} \eta n_1^{1/2}$G.  realistic cases.

\end{abstract}

\keywords{gamma rays: bursts- hydrodynamics-relativity - shock waves}

\section{Introduction}
 Understanding the phenomenon of GRBs is one of the important problems of
recent astrophysics. Fortunately, following the discovery of a
cosmological redshift in the May 08, 1997 event, it is certain now that
some or all of them could be of cosmological origin Metzger et al. (1997).
Whether cosmological or galactic, GRB phenomenon is broadly understood in
terms of a standard model developed by Cavallo \& Rees (1978), Goodman
(1986), Paczynski (1986), Eichler et al. (1989), Shemi \& Piran (1990).
Nonetheless, as far as the origin of the complex nonthermal observed GRB
spectra are concerned, an important development took place with the work
of Rees \& Meszaros (1992) and Meszaros \& Rees (1993) suggesting that the
cosmic fireballs (FB) with an optimal amount of baryonic pollution, $\eta
\sim 10^2 -10^3$, could explain such spectra, where $\eta \equiv E/Mc^2$,
 where $E$ is the  energy and $M$ is the mass of the baryons entrained in
the original fireball (FB).  Meszaros \& Rees suggested that as the baryon
polluted FB deposits half of its original momentum onto the ambient
medium, presumably, the bare interstellar medium (ISM), at $r=r_{\rm d}$,
the so-called deceleration radius, the blast wave becomes very strong and
radiates part of its energy. For further appreciation of this paper it
would be appropriate to crudely visualize the geometry associated with the
blast wave in terms of a 1-D simple diagram (Mitra 1998, henceforth M98).
Here region (1) is the ambient ISM, the lab frame, $S_1$ is the forward
shock moving ahead of the contact discontinuity $S$, the location of the
original FB boundary.  Region (2) represents the (forward) shocked fluid
and it is this region which is the site for the particle acceleration and
gamma ray production in this standard model. The region (4) is the
unperurbed FB and (3) is the part of the FB compressed by the reverse
shock front $S_2$. It was shown in M98 that, in the context of this
standard model, the reverse shock plays an insignificant role in the
overall energy balance and may be neglected for dynamical purposes.

The gamma rays are likely to be produced either by a synchrotron process
or a self-synchrotron-Compton process occuring near the region (2) and the
most crucial factor for the success of such processes is the value of the
comoving magnetic field $B_2'=B_{\rm sh}$. Here prime denotes respective
comoving quantities, i.e. respective {\em proper values},  whereas `*'
denotes quantities measured in the rest frame of the forward shock $S_1$.
The question of probable generation of a magnetic field in a relativistic
(or nonrelativistic) shock is a poorly understood topic, and, practically,
most of the authors have therefore been compelled to use an equiparition
argument to estimate the same (Meszaros \& Rees 1993, Cheng \& Wei 1996,
Vietri 1995, Waxman 1995):
\begin{equation}
{B_{\rm sh}^2\over 8 \pi} \sim \eta^2 n_1 m c^2
\end{equation}
But  by recalling the basic definition of
\begin{equation}
r_{\rm d} \approx \left({3 E \over 2\pi c^2 \eta^2 m n_1}\right)^{1/3}
\approx 7\times 10^{15} E_{51}^{1/3} \eta_3^{-2/3} n_1^{-2/3}~{\rm cm}
\end{equation}
where $\eta_3= 10^{-3} \eta$ and $E_{51}= E/10^{51}$erg s$^{-1}$, it can be
easily verified that the energy density shown on the R.H.S. of eq.(1)
directly corresponds to the region (4), i.e, the unperturbed FB, having a
proper density (M98)
\begin{equation}
n_4'= {E(\gamma_{\rm F}/ \eta) \over 4\pi r^3 c^2 m} \approx 5\times 10^7
E_{51} r_{15}^{-3} (\gamma_{\rm F}/\eta)~ {\rm cm}^{-3},
\end{equation}
i.e., actually,
\begin{equation}
{B_4'^2\over 8 \pi} \approx \gamma_{\rm F}^2 n_1 m c^2
\end{equation}
 Here $\gamma_{\rm F}$ is the bulk Lorentz factor (LF) of $S_1$ in the lab
frame (1), $m$ is the mass of a proton, and $n_1$ is the particle number
density of the ambient medium in units of  1$cm^{-3}$. Note that for the
region (1) the comoving frame coincides with the lab frame and at
$r=r_{\rm d}$, we will have $\gamma_{\rm F} \approx \eta/2$ in the
Meszaros \& Rees scenario.  On the other hand, we need to apply the
equipartition argument in the downstream of the shock, i.e., in the
region, which is expected to be turbulent, and which is, in any case, the
site for the particle acceleration. And, it will be seen in the next
section that, equipartition argument yields approximately the same value
of $B_2'$ in the shocked fluid, and this is important, because, energy
density, in itself, is not a Lorentz invariant quantity.

Nevertheless, the question we want to pose here is how justified is this
assumption of equipartition in the context of GRBs and whether by adopting
this brute assumption we are running into conflict with some well
established feature of relativistic collisionless shocks. As was stressed
in Mitra (1996), equipartition, as a general physical concept may be found
to be valid in steady-state situations like the ISM where the plasma
interacts with the particles and currents over astronomically significant
time scales. As to dynamic situations, there are hints that many young
supernova remnants are endowded with freshly generated magnetic fields
which are considerably higher than the bare ISM values $\sim 3\times
10^{-6}$G. Nonetheless, even in such cases, the age of the supernova could
be thousands of years and the enhanced magnetic field is usually much
smaller than what is obtained by naive equipartition arguments.  In fact,
it was clearly anticipated by Meszaros, Rees, \& Papathanssiou (1994) that
the equipartition argument can at best serve as a broad guide to determine
the actual value of $B_{\rm sh}$, and accordingly, they introduced a {\em
completely free parameter}, $\lambda \le 1$,  tagged onto the naively
obtained value of $B_{\rm sh}^{\rm eq}$:
\begin{equation}
B_{\rm sh} \sim 4\times 10^2 n_1^{1/2} \eta_3 \lambda^{1/2} ~{\rm G}
\end{equation}
Further, it could be possible to apply the basic equipartition idea at
$t=0$ to the initial FB, and then evaluate the value of the instantaneous
$B_{\rm FB}$ or $B_{\rm sh}$ by using the flux-freezing condition. And
again, in this case, if we symbolize our ignorance through the free
parameter, $\xi \le 1$, it follows that (Meszaros, Rees \& Papathanassiou
1994)
\begin{equation}
B_{\rm sh} \sim 0.4 \xi^{1/2} E_{51}^{-1/6} n_1^{2/3} \eta_3^2 ~{\rm G}
\end{equation}
Thus, in this paper, we would attempt to invoke a known feature of relativistic
collisionless MHD shocks to obtain physically significant upper limits on
$B_{\rm sh}$.

\section{Shock Dynamics}
Following the work on relativistic strong shock jump conditions by Taub
(1949) and Blandford \& McKee (1976) it was discussed in M98 that
the ratio of the comoving particle densities in the region (1) and (2) is
\begin{equation}
{n_2'\over n_1'} = {\Gamma_2 \gamma_{12} +1 \over \Gamma_2 -1}
\end{equation}
where $\Gamma_2 \approx 4/3$ is the effective ratio of the specific heat
of (2) and $\gamma_{12}$ is the bulk LF of (2) with respect to (1).
Therefore, we have
\begin{equation}
n_2' \approx (4 \gamma_{12} +3) n_1' \approx 4 \gamma_{12} n_1 \approx
2\sqrt{2} \gamma_{\rm F} n_1
\end{equation}
where we have used the fact  that the maximum value of the LF of the (forward)
shocked fluid with respect to the FB is only $\sqrt{2}$ (M98) :
\begin{equation}
{\gamma_{\rm F}\over \sqrt{2}} \le \gamma_{12} \le \gamma_{\rm F}
\end{equation}
The Fig.1 actually represents a lab-frame vision where the shock front
$S_1$ is moving. However, shock dynamics is often better studied in the
rest frame of $S_1$ (*) where, the region (1) (upstream) would be seen to
gushing in with a large bulk LF, $\gamma_{1*}=\gamma_*$. On the other
hand, the region (2) (downstream) would be seen to run away with small LF
$\gamma_{2*} \lesssim \sqrt{2}$ (M98). The shock jump conditions also show
that the proper relativistic internal energy density of the shocked fluid is
\begin{equation}
e_2' \approx  \gamma_{12} n_2' c^2 \approx   m n_1 \eta^2 c^2
\end{equation}
which shows that an equipartition argument in region (2) approximately leads to
a value of $B_{\rm sh}$ as is given by eq. (1):
\begin{equation}
B_2'=B_{\rm sh} \sim 150 \eta_3^{1/2}~n_1^{1/2} ~{\rm G}
\end{equation}
If the plasma is supposed to have infinite conductivity, all comoving
electric fields must vanish. Then, for a parallel shock where the embedded
magnetic field is parallel to the flow direction (x-axis) one obtains the
following simple shock jump condition (de Hoffman \& Teller 1950,
henceforth HT50):
\begin{equation}
B_{2 \rm x}' =B_{1 \rm x}'; \qquad B_{\rm y}'=B_{\rm z}'=0
\end{equation}
Further, since under Lorentz transformation, the parallel component of
magnetic field remains unchanged,  for a parallel shock, we find
\begin{equation}
B_{1 \rm x}'=B_{1 \rm x *} =B_{2 \rm x}'=B_{2 \rm x *}= B_{\rm sh}
\end{equation}

\subsection{Perpendicular shock}
For a perpendicular shock, the magnetic field changes under Lorentz
transformation and accordingly shock jump conditions are different from
their parallel counterpart. If the magnetic field lies in the y-direction
($B_{\rm x}=B_{\rm z}=0$), Lorentz transformation leads to (HT50)
\begin{equation}
B_{1\rm y}^*=\gamma_* B_{1\rm y}'; \qquad B_{2\rm y}^*=\gamma_{2*} B_{2\rm
y}',
\end{equation}
where $\gamma_{2*} \lesssim \sqrt{2}$ is the LF of the downstream fluid in
the shock frame (M98), and all other components of magnetic field in any
other frame becomes zero.  Further for all transverse cases, the frozen-in
magnetic field line densities become proportional to the respective
comoving particle densities (HT50):
\begin{equation}
{B_{2\rm y}'\over B_{1\rm y}'} = {n_2'\over n_1'}
\end{equation}
Then eq. (14) and (15) together imply
\begin{equation}
{B_{2\rm y}^* \over B_{1\rm y}^*} ={\gamma_{2*} \over \gamma_*} {n_2'\over
n_1'} \approx 4
\end{equation}
Thus, for the transverse case too, the magnetic field jump condition looks
similar to the corresponding non-relativistic case if quantities are
measured in the shock frame. However, actually, there is a strong
enhancement of the magnetic field in the downstream and this becomes clear
if we write the above equation in terms of respective proper fields by
using eq. (8):
\begin{equation}
B_2' \approx 4 \gamma_{12} B_1' \approx 4 \gamma_* B_1'\approx 
2\sqrt{2} \gamma_{\rm F} B_1' 
\approx \sqrt{2} \eta B_1'
\end{equation}

\section{Relativistic MHD shocks}
Although, the micro-physics of relativistic shocks, in particular, collisionless
MHD shocks is poorly understood, atleast in comparison to their
non-relativistic counterparts, there have been considerable amount of
observational, numerical as well as analytical studies of relativistic
collisionless shocks. Although, such studies have largely been focussed
for the perpendicular or transverse shocks, presumably, to harness the
shock drift mechanism inherent in the $\vec{v} \times \vec{B}$ term, we
will see that some fundamental aspect of such studies may be extendable to
the the parallel cases too. To be precise, it is found that the following parameter
\begin{equation}
\sigma\equiv {v_{1*}(B_{{1\rm y}^*}^2/4\pi) \over v_{1*}(2m n_{1*} \gamma_{1*}
c^2)},
\end{equation}
where $v_{1*}$ is the speed of the upstream fluid in the shock frame, is
very important for studying the structure and accelerating properties of
the shock (Kennel \& Coroniti 1984a, 1984b, Alsop \& Arons 1988, Hoshino
et al. 1992). Physically $\sigma$ is the ratio of the upstream Poynting
energy flux and the the particle energy density as measured in the {\em
shock frame}. Stated  this way, it may appear that $\sigma$ may not have
any relevance for the parallel case.  However, by noting that $B_{1\rm
y}^* =\gamma_* B_{1\rm y}'=\gamma_* B_1'$ and that the shock frame
upstream particle density $n_{1*}=\gamma_* n_1$, one can reformulate eq. (18)
as
\begin{equation}
\sigma= {B_1'^2/8 \pi \over m n_1 c^2}
\end{equation}
One can clearly see that the above definition of $\sigma$ is Lorentz
covariant and of a general nature :
\begin{equation}
\sigma={upstream ~proper~ magnetic~ energy~ density \over upstream~rest~
mass~ energy~ density}
\end{equation}
Therefore, the foregoing definition of $\sigma$ becomes relevant for all
cases.  Also, note that, in contrast to non-relativistic shocks, for all
practical purposes, ultrarelativistic shocks are bound to be tranverse
ones because, the shock may be considered to be parallel only if
$\theta_{\rm B_1} <\gamma_*^{-1}$, where, $\theta_{\rm B_1}$ is the angle
between the upstream (proper) magnetic field and the flow direction as
measured in the upstream proper frame (Begelman \& Kirk 1990, Hoshino et
al. 1992). In the present case, the upstream proper frame is the ISM and
there is no reason that the ISM field (which is in any case distorted over
certain length scale) should be practically perfectly aligned with the GRB
flow direction. Therefore, irrespective of the Lorentz covariant
definition of $\sigma$, the idealized discussion of a perpendicular
relativistic shock becomes quite relevant in the present case.

Note that, in case one is performing numerical experiments to study the
formation of shocks out of relativistic flows, one must consider a wide
range of $\sigma$ including those $>1$ (Langdon, Arons, \& Max 1988). But,
it should be realized that a value of $\sigma >1$ necessarily means that
the upstream flow has enough internal energy to support current systems
{\em whose energy density exceeds the rest mass energy density}.  In other
words, such numerical experiments {\em correspond to an upstream which is
relativistically, ``hot''}. For most of the realistic astrophysical
situations, theoretical and observational arguments suggest that the value
of $\sigma \ll 1$ (Piddington 1957, Rees \& Gunn 1974, Kennel \& Coroniti
1984a,b., Hoshino et al. 1992). And, in any case, for the cold upstream
region, which is the  appropriate for the present case and most
astrophysical situations, an absolute upper limit is $\sigma< 1$. If we,
somewhat naively, use this upper limit on $\sigma$ for a parallel shock,
we obtain
\begin{equation}
{B_1'^2\over 8 \pi} = m n_1 c^2
\end{equation}
And this leads to an absolute upper limit for a supposed parallel
relativistic shock is
\begin{equation}
B_{\rm sh}^\parallel =B_2'= B_1' \approx 0.2 n_1^{1/2} ~{\rm G}
\end{equation}
The basic reason that for the parallel case, we are considering the above
mentioned upper limit on $\sigma$ is that, we are not aware of any
theoretical or numerical study of relativistic parallel shocks which may
tighten this constraint further. Coming back to the fairly well studied
case of perpendicular relativistic shocks, we may adopt a generous upper
limit on the value of $\sigma \approx 0.002$ by simply adopting the value
corresponding to that of Crab (Rees \& Gunn 1974, Kennel \& Coroniti
1984a,b):
\begin{equation}
B_{1 \rm crab}'\approx 9 \times 10^{-3} n_1^{1/2}~{\rm G}
\end{equation}
Or,
\begin{equation}
B_{\rm sh}=B_2' \approx 4 \gamma_{12} B_{1 \rm crab}'\approx 15~ \eta_3
n_1^{1/2}~{\rm G}
\end{equation}
Given this generous upper-limit on $\sigma$, depending on the actual
obliquity of a given shock, the actual value of $B_{\rm sh}$ will vary
between what is shown by eq. (22) and (24) with more likelyhood of
assuming the latter value.  This condition that we must have $\sigma \ll
1$ (atleast for perpendicular shocks) in order to have a strong and
accelerating shock is somewhat akin to the condition that for
non-relativistic diffusive shock acceleration the Alfven Mach number
$M_{\rm A} \gg1$ (Begelman \& Kirk 1990). This condition means that the
Alfven speed must be $<<$ than the flow speed to ensure that MHD
scattering centers scatter the test particles vigourously and
isotropically. This condition also implies that the magnetic energy
density is negligible compared to the kinetic energy flux:
\begin{equation}
{B_2'^2 \over 8 \pi} << {1\over 2} m n_2' v_2^2
\end{equation}
For nonrelativistic perpendicular shocks too, similar conditions are
necessary to ensure that the shock is ``strong'' (Kundt \& Krotschek 1982,
Leroy et al. 1982, Appl \& Camenzind 1988).

\section{Discussion}
Having obtained this generous upper-limit let us now ponder how justified
we are in adopting a value of $\sigma$ appropriate for Crab. Remember that
the Crab shock is practically a standing shock and the upstream region is
not the bare ISM. On the other hand, the {\em upstream comprises the
plasma ejected by the Crab pulsar during its life time of nearly thousand
years}.  It is very much likely that, in the past, the value of $\sigma$
for Crab was much lower, and the present value has been slowly built up
over these thousand years. In contrast the case of the GRB blast wave
ploughing through the bare ISM is quite different, and, it is highly
improbable that, within a lab frame time scale of few seconds or less, the
upstream medium ahead of the shock front (as seen by in the lab frame) can
raise its magnetic field from a value of $\sim 3 \times 10^{-6}$G to $\sim
9\times 10^{-3}$G.  We feel that, instead, the following scenario is more
plausible: The value of $B_1'$ probably remains close to its unperturbed
value $\sim 3\times 10^{-6}$G; however the shock could be near
perpendicular resulting in a value of $B_{\rm sh} \sim 4 \gamma_{12}
B_1'\sim \sqrt{2} \eta B_1' \sim 4 10^{-3}
\eta_3$G. It is also probable, there may not be any stable shock formation
at all invalidating the rigid relations between $B_1'$ and $B_2'$ employed
so far, and on the other hand there may be instantaneous spikes in the
downstream magnetic field (Langdon, Arons, \& Max 1988) :
\begin{equation}
{B_{2\rm max}^* \over B_{\rm sh}^*} \approx \left[ 1+ (2/\sigma) \right]^{1/2}
\end{equation}
and which may erratically raise the shocked field to a $B_{\rm max} \sim
0.1 -1$G.  However, formation of a shock-like discontinuity, either steady
or fluctuating requires that the ambient medium should be such that the
{\em leading particles} of the FB, i.e., the piston driving the shock,
either individually or collectively  impart significant amount of their
momentum on the ambient medium. We endeavoured to examine this critical
but usually overlooked problem in M96. Noting the similarity between the
present problem and the one involving propagation of high energy cosmic
rays in the ISM, and also recalling that phenomenologically and
observationally obtained parameter, the {\em spatial diffusion
coefficient} describes the entire collective interaction of the cosmic
rays and the ISM, we found, that, it is implausible that the FB can
produce any thing akin to a GRB blast wave in the ISM (M96). This is so
for the simple reason that the BF-ISM interaction time scale estimated in
this way could be as large as $\sim 10^7$s !

Probably, we may also examine here whether, the electrons or the left over
pairs of the FB, rather than the protons, might carry out the job of
imparting energy onto the ambient medium (B. Paczynski, private
communication).  For a given saturation bulk LF of the FB ($\gamma_{\rm F}
\approx \eta/2$), the electrons or positrons will be less energetic by a
factor of $\sim 2000$. And, if we are assuming the Bohm limit of the
diffusion coefficient, the deflection length as well as the lab frame
deflection time will accordingly be smaller by a factor of $\sim 2000$. If
the numer of leptons in the FB could overwhelm the protons by a factor of
$\sim 2000$, there would have been equal amount of energy residing in
protons and leptons, and, the foregoing reduction in time scale would have
really meant that the FB-ISM interaction time would be reduced by the same
factor. Unfortunately, this is not the case, the number of leptons per
proton at the saturation stage of the optimally baryon polluted FB is
indeed $\gtrsim 1$. This means that the eventual FB-ISM interaction time
scale should be what was obtained in M96, i.e.,
\begin{equation}
t_i \sim 3 \times 10^4 \eta_3 ~{\rm s}
\end{equation}
unless it is found that the lepton-ion energy transfer time scale is
smaller than this above time scale. It may be probable that at best there
may be soliton like discontinuites. At any rate, we do not think, we have
the final answer of such questions at this moment, and we realize that, we
are simply attempting to understand various aspects subject to  our
(present author's) available knowledge.

The upshot of this discussion is that the original GRB is likely to be
produced either by internal collisions within the FB as has long been
suspected (Paczynski \& Xu, Rees \& Meszaros 1994, Papathanssiou \&
Meszaros 1996) or if the original FB is propagating within a medium which
has a modest baryonic mass but is dense enough to absorb the FB momentum
either by binary collisions or by collisionless MHD process. Finally,
reverting back to the original mandate of this paper, in the case of a
supposed GRB blast wave propagating in the ISM and even ignoring the
disturbing possibility that no shock like discontinuity may be formed on
the GRB time scale, we feel that, the maximum value of the magnetic field
in the shocked fluid may not exceed $\sim 1 $G.  We would like to
emphasize here the fact that this conclusion does not at all imply that
the FB can not excite a blast wave (initially relativistic and then
non-relativistic) on a much larger time scale of days or weeks and which is
necessary for explaining the GRB after glow in various low energy
brackets.


\centerline{Figure Caption}

Figure 1: Sketch of the FB-shock configurations, for details, see text.

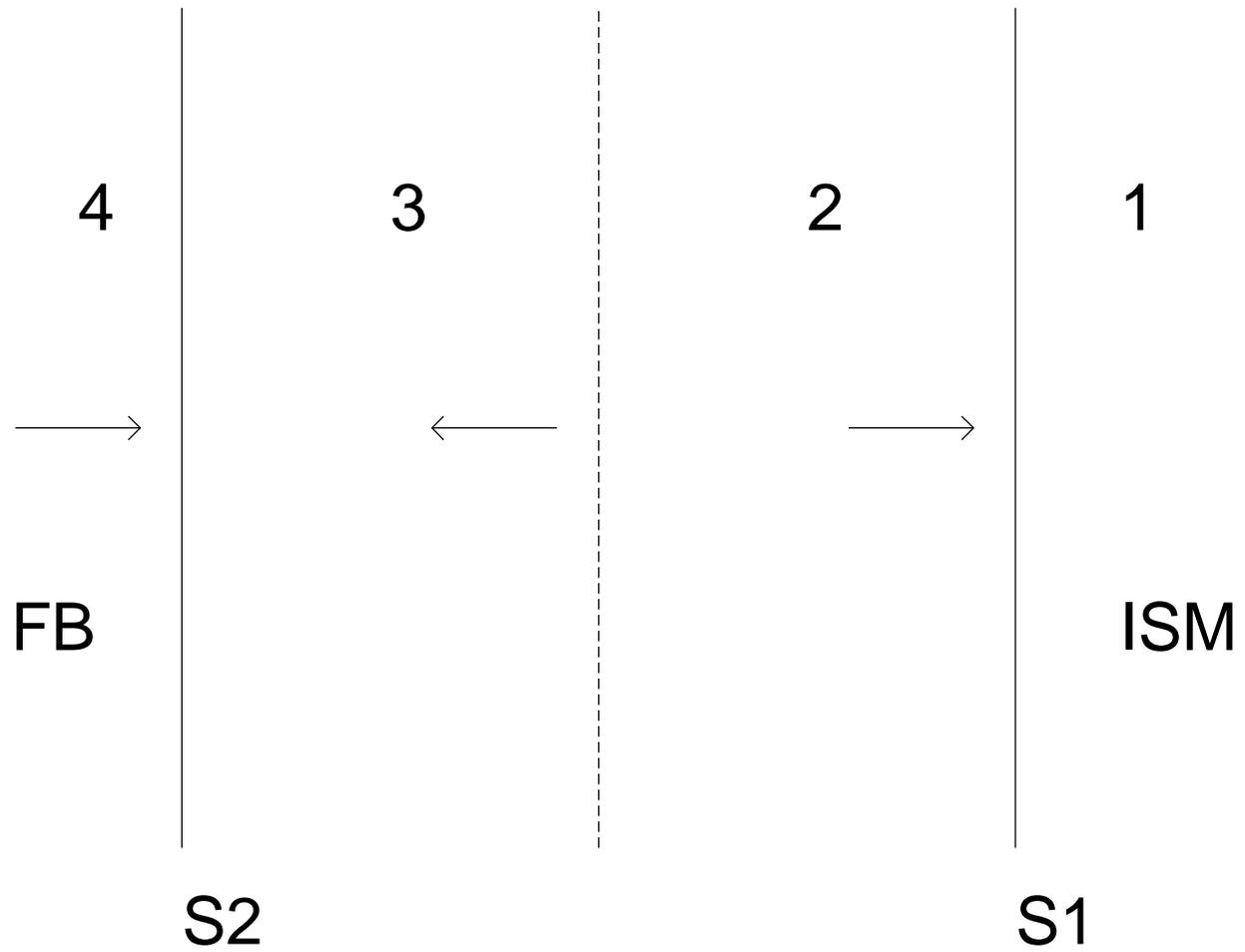
\end{document}